\documentclass[nofootinbib,preprint,preprintnumbers,a4paper,10pt]{revtex4}
\usepackage{multirow}
\usepackage{textcomp}
\usepackage{amsmath,graphicx,color,epsfig}

\usepackage{footnote}
\usepackage{ulem}
\usepackage{booktabs}
\usepackage{array}
\usepackage{amssymb}
\usepackage{subfigure}
\usepackage{longtable}
\usepackage{verbatim}
\usepackage{amsfonts}
\usepackage{hyperref}
\usepackage{cancel}

\begin{document}

\title{Topological diagram analysis of $\mathcal{B}_{c\overline 3}\to \mathcal{B}_{10}M$ decays \\ in the $SU(3)_F$ limit and beyond}

\author{Di Wang$^{1}$}\email{wangdi@hunnu.edu.cn}

\address{%
$^1$Department of Physics, Hunan Normal University, Changsha 410081, China
}

\begin{abstract}
Charm baryon decay plays an important role in studying non-perturbative baryonic transitions.
Compared to other hadron multiplets, the flavor symmetry of baryon decuplet is more simple and attractive.
In this work, we study the topological amplitudes of charmed baryon decays into decuplet baryon in the flavor symmetry and the linear $SU(3)_F$ breaking.
It is found most of topological diagrams are suppressed by the K\"orner-Pati-Woo theorem in the $SU(3)_F$ limit.
Only two independent amplitudes contributing to the $\mathcal{B}_{c\overline 3}\to \mathcal{B}_{10}M$ decays, with one dominating the branching fractions.
The Lee-Yang parameters of all $\mathcal{B}_{c\overline 3}\to \mathcal{B}_{10}M$ modes are the same in the $SU(3)_F$ limit, and there are only four possible values for the CP asymmetries.
After including the first-order $SU(3)_F$ breaking effects, the $\Xi^+_c\to \Sigma^{* +}\overline K^0$ and $\Xi^+_c\to \Xi^{* 0}\pi^+$ decays have non-zero branching fractions.
The number of free parameter contributing to the $\mathcal{B}_{c\overline 3}\to \mathcal{B}_{10}M$ decays in the linear $SU(3)_F$ breaking is smaller than the available data.
The $SU(3)_F$ breaking part of the quark loop diagram can be extracted by global fitting of branching fractions, which could help us understand the CP violation in charm sector.
Additionally, some new isospin equations are proposed to test the K\"orner-Pati-Woo theorem.
\end{abstract}

\maketitle

\section{Introduction}
Charmed baryon weak decay plays an important role in studying the weak and strong interactions and the nature of $b$-baryons.
In experimental aspects, most efforts are devoted to investigate the charmed baryon decays into octet baryons \cite{PDG}, while
decay modes involving decuplet baryons are not richly observed.
Due to the total symmetric flavor wavefunction of decuplet, the theoretical analysis of decays into decuplet baryons is easier than those involving octet baryons.
For example, the K\"orner-Pati-Woo (KPW) theorem \cite{Korner:1970xq,Pati:1970fg} was applied to the $\mathcal{B}_{c\overline 3}\to \mathcal{B}_{10}M$ decays in Ref.~\cite{Geng:2019awr}.
It showed that most of decay amplitudes are suppressed in the $SU(3)_F$ limit, and only one amplitude dominates the $\mathcal{B}_{c\overline 3}\to \mathcal{B}_{10}M$ decays.
The $\mathcal{B}_{c\overline 3}\to \mathcal{B}_{10}M$ decays provide an ideal platform for studying the weak and strong interactions at charm scale.

Topological diagram is an intuitive tool to analyze the dynamics of heavy hadron decays and has been employed in studies of charmed baryon decays \cite{Wang:2024ztg,Groote:2021pxt,Zhao:2018mov,He:2018joe,Zhong:2024qqs,Hsiao:2021nsc,Kohara:1991ug,Chau:1995gk,Zhong:2024zme}.
It provides a framework for both model-dependent data analysis and model calculations.
The topological diagrams of the $\mathcal{B}_{c\overline 3}\to \mathcal{B}_{10}M$ decays have been studied in Ref.~\cite{Hsiao:2020iwc}.
However, the quark loop diagrams were not included, and thus the CP asymmetries were not analyzed.
Besides, the $SU(3)_F$ breaking effects were not formulated in Ref.~\cite{Hsiao:2020iwc}.
Thereby, it is significative to revisit the topological amplitudes of $\mathcal{B}_{c\overline 3}\to \mathcal{B}_{10}M$ decays in the flavor symmetry and beyond.

In this work, we first study the topological amplitudes of the $\mathcal{B}_{c\overline 3}\to \mathcal{B}_{10}M$ decays in the $SU(3)_F$ symmetry.
It is found there are only two independent amplitudes contributing to the $\mathcal{B}_{c\overline 3}\to \mathcal{B}_{10}M$ decays and one of them is dominated in the branching fractions.
The Lee-Yang parameters of all $\mathcal{B}_{c\overline 3}\to \mathcal{B}_{10}M$ modes are identical,
and there are only four possible values for the CP asymmetries in these decays.
We then investigate the $\mathcal{B}_{c\overline 3}\to \mathcal{B}_{10}M$ decays in the linear $SU(3)_F$ breaking.
We find the number of free parameter contributing to the $\mathcal{B}_{c\overline 3}\to \mathcal{B}_{10}M$ decays is fewer than the available data.
The $SU(3)_F$ breaking part of quark loop diagram can be extracted by global fitting, aiding our understanding of CP violation in the charm sector.
The branching fractions of $\Xi^+_c\to \Sigma^{* +}\overline K^0$ and $\Xi^+_c\to \Xi^{* 0}\pi^+$ are non-zero after considering the first-order $SU(3)_F$ breaking effects.
And some new isospin equations are found to test the K\"orner-Pati-Woo theorem and isospin symmetry.

The rest of this paper is structured as follows.
In Sec.~\ref{sy}, we perform a topological analysis of the $\mathcal{B}_{c\overline 3}\to \mathcal{B}_{10}M$ decays in the $SU(3)_F$ limit.
The topological amplitudes of the $\mathcal{B}_{c\overline 3}\to \mathcal{B}_{10}M$ decays in the linear $SU(3)_F$ breaking are investigated in Sec.~\ref{be}.
And Sec.~\ref{sum} is a short summary.

\section{Topological amplitudes in the $SU(3)_F$ limit}\label{sy}
In this section, we first present the topological diagrams of charmed baryon decays into a decuplet baryon and a light meson.
Then we apply the KPW theorem to reduce decay amplitudes.
The effective Hamiltonian in charm quark decay in the SM can be written as \cite{Buchalla:1995vs}
 \begin{equation}\label{hsm}
 \mathcal H_{\rm eff}={G_F\over \sqrt 2}
 \left[\sum_{q=d,s}V_{cq_1}^*V_{uq_2}\left(\sum_{q=1}^2C_i(\mu)\mathcal{O}_i(\mu)\right)
 -V_{cb}^*V_{ub}\left(\sum_{i=3}^6C_i(\mu)\mathcal{O}_i(\mu)+C_{8g}(\mu)\mathcal{O}_{8g}(\mu)\right)\right],
 \end{equation}
 where $G_F$ is the Fermi coupling constant, $C_{i}$ is the Wilson coefficients of operator $\mathcal{O}_i$.
The magnetic-penguin contributions can be included into the Wilson coefficients for the penguin operators following the substitutions
\cite{Beneke:2003zv,Beneke:2000ry,Beneke:1999br}
\begin{eqnarray}
C_{3,5}(\mu)\to& C_{3,5}(\mu) + \frac{\alpha_s(\mu)}{8\pi N_c}
\frac{2m_c^2}{\langle l^2\rangle}C_{8g}^{\rm eff}(\mu),\qquad
C_{4,6}(\mu)\to& C_{4,6}(\mu) - \frac{\alpha_s(\mu)}{8\pi }
\frac{2m_c^2}{\langle l^2\rangle}C_{8g}^{\rm eff}(\mu),\label{mag}
\end{eqnarray}
with the effective Wilson coefficient $C_{8g}^{\rm eff}=C_{8g}+C_5$.
In the $SU(3)$ picture, the weak Hamiltonian of charm decay can be written as \cite{Wang:2020gmn}
 \begin{equation}\label{h}
 \mathcal H_{\rm eff}= \sum_p \sum_{i,j,k=1}^3 (H^{(p)})_{ij}^{k}\mathcal{O}_{ij}^{(p)k},
 \end{equation}
in which
\begin{equation}
\mathcal{O}_{ij}^{(p)k} = \frac{G_F}{\sqrt{2}} \sum_{\rm color} \sum_{\rm current}C_p(\overline q_iq_k)(\overline q_jc),
\end{equation}
and superscript $p=0,1$ denote the tree and penguin operators.
$(H^{(p)})_{ij}^k$ can be obtained from the map $(\bar uq_1)(\bar q_2c)\rightarrow V^*_{cq_2}V_{uq_1}$ in current-current operators and $(\bar qq)(\bar uc)\rightarrow -V^*_{cb}V_{ub}$ in penguin operators according to Eq.~\eqref{hsm}.
The non-zero $(H^{(0)})_{ij}^k$ induced by tree operators include
\begin{align}\label{ckm1}
 &(H^{(0)})_{13}^2 = V_{cs}^*V_{ud},  \qquad (H^{(0)})^{2}_{12}=V_{cd}^*V_{ud},\qquad (H^{(0)})^{3}_{13}= V_{cs}^*V_{us}, \qquad (H^{(0)})^{3}_{12}=V_{cd}^*V_{us}.
\end{align}
 The non-zero $(H^{(1)})_{ij}^k$ induced by penguin operators include
\begin{align}\label{ckm2}
 &(H^{(1)})_{11}^1 = -V_{cb}^*V_{ub}, \qquad (H^{(1)})_{21}^2=-V_{cb}^*V_{ub}, \qquad (H^{(1)})_{31}^3=-V_{cb}^*V_{ub}.
\end{align}

In the flavor $SU(3)$ symmetry, the charmed anti-triplet baryon is expressed as
\begin{eqnarray}
 \mathcal{B}_{c\overline 3}=  \left( \begin{array}{ccc}
   0   & \Lambda_c^+  & \Xi_c^+ \\
    -\Lambda_c^+ &   0   & \Xi_c^0 \\
    -\Xi_c^+ & -\Xi_c^0 & 0 \\
  \end{array}\right).
\end{eqnarray}
The light pseudoscalar nonet meson is
\begin{eqnarray}
 M=  \left( \begin{array}{ccc}
   \frac{1}{\sqrt 2} \pi^0+  \frac{1}{\sqrt 6} \eta_8    & \pi^+  & K^+ \\
    \pi^- &   - \frac{1}{\sqrt 2} \pi^0+ \frac{1}{\sqrt 6} \eta_8   & K^0 \\
    K^- & \overline K^0 & -\sqrt{2/3}\eta_8 \\
  \end{array}\right) +  \frac{1}{\sqrt 3} \left( \begin{array}{ccc}
   \eta_1    & 0  & 0 \\
    0 &  \eta_1   & 0 \\
   0 & 0 & \eta_1 \\
  \end{array}\right).
\end{eqnarray}
The $\eta_8$ and $\eta_1$ are not mass eigenstates.
The mass eigenstates $\eta$ and $\eta^\prime$ are mixing of $\eta_8$ and $\eta_1$,
\begin{eqnarray}
\left( \begin{array}{ccc}
\eta\\
\eta^\prime
\end{array}
\right)
=
\left(
\begin{array}{cc}
\cos\xi  &  -\sin\xi\\
\sin\xi  &  \cos\xi
\end{array}
\right)\left(
\begin{array}{c}
\eta_8\\
\eta_1
\end{array}\right).
\end{eqnarray}
The light baryon decuplet is given as
\begin{align}
 &\Delta^{++}  = \mathcal{B}_{10}^{111},  \quad \Delta^{-}=\mathcal{B}_{10}^{222},\quad  \Omega^-=\mathcal{B}_{10}^{333}, \quad\Sigma^{*0}=\frac{1}{\sqrt{6}}(\mathcal{B}_{10}^{123} + \mathcal{B}_{10}^{132} +\mathcal{B}_{10}^{213}+ \mathcal{B}_{10}^{231} + \mathcal{B}_{10}^{312} +\mathcal{B}_{10}^{321}),\nonumber\\
 & \Delta^{+} = \frac{1}{\sqrt{3}}(\mathcal{B}_{10}^{112} + \mathcal{B}_{10}^{121} + \mathcal{B}_{10}^{211}),\quad  \Delta^{0} = \frac{1}{\sqrt{3}}(\mathcal{B}_{10}^{122} + \mathcal{B}_{10}^{212} + \mathcal{B}_{10}^{221}),\quad \Sigma^{*+}= \frac{1}{\sqrt{3}}(\mathcal{B}_{10}^{113} +\mathcal{B}_{10}^{131} +\mathcal{B}_{10}^{311}), \nonumber\\
&\Sigma^{*-}=\frac{1}{\sqrt{3}} (\mathcal{B}_{10}^{223} +\mathcal{B}_{10}^{232} +\mathcal{B}_{10}^{322}), \quad
 \Xi^{*0}=\frac{1}{\sqrt{3}}(\mathcal{B}_{10}^{133} +\mathcal{B}_{10}^{313} +\mathcal{B}_{10}^{331}),\quad
 \Xi^{*-}=\frac{1}{\sqrt{3}}(\mathcal{B}_{10}^{233} + \mathcal{B}_{10}^{323}+\mathcal{B}_{10}^{332}).
\end{align}
The decay amplitude of $\mathcal{B}_{c\overline 3}\to \mathcal{B}_{10}M$ mode can be constructed by
\begin{align}\label{am1}
  \mathcal{A}(\mathcal{B}_{c\overline 3}\to \mathcal{B}_{10}M)& = E^\prime(\mathcal{B}_{c\overline3})_{ij}H^j_{kl}M^i_m \mathcal{B}_{10}^{klm}+C^{\prime}(\mathcal{B}_{c\overline3})_{ij}H^k_{lm}M^j_k \mathcal{B}_{10}^{ilm}
  +E_M(\mathcal{B}_{c\overline3})_{ij}H^j_{kl}M^l_m \mathcal{B}_{10}^{ikm}\nonumber\\
&+  E_B(\mathcal{B}_{c\overline3})_{ij}H^j_{kl}M^k_m \mathcal{B}_{10}^{ilm} +E_S(\mathcal{B}_{c\overline3})_{ij}H^j_{kl}M^m_m \mathcal{B}_{10}^{ikl}
  +L(\mathcal{B}_{c\overline3})_{ij}H^l_{kl}M^j_m \mathcal{B}_{10}^{ikm}
\nonumber\\
&+L_1(\mathcal{B}_{c\overline3})_{ij}H^l_{lk}M^j_m \mathcal{B}_{10}^{ikm}.
\end{align}
The completeness of Eq.~\eqref{am1} can be proven by permutation.
The sum of decay amplitudes contributing to $\mathcal{B}_{c\overline 3}\to \mathcal{B}_{10}M$ and $\mathcal{B}_{c 6}\to \mathcal{B}_{10}M$ modes is $N_{\overline 3}+N_6=A^5_5/A_3^3=20$, and the deference of decay amplitudes contributing to $\mathcal{B}_{c6}\to \mathcal{B}_{10}M$ and $\mathcal{B}_{c\overline 3}\to \mathcal{B}_{10}M$ modes is $N_6-N_{\overline 3}=A_3^3=6$.
Solving these equations gives $N_6=13$, $N_{\overline 3}=7$.

\begin{figure}
  \centering
  \includegraphics[width=15cm]{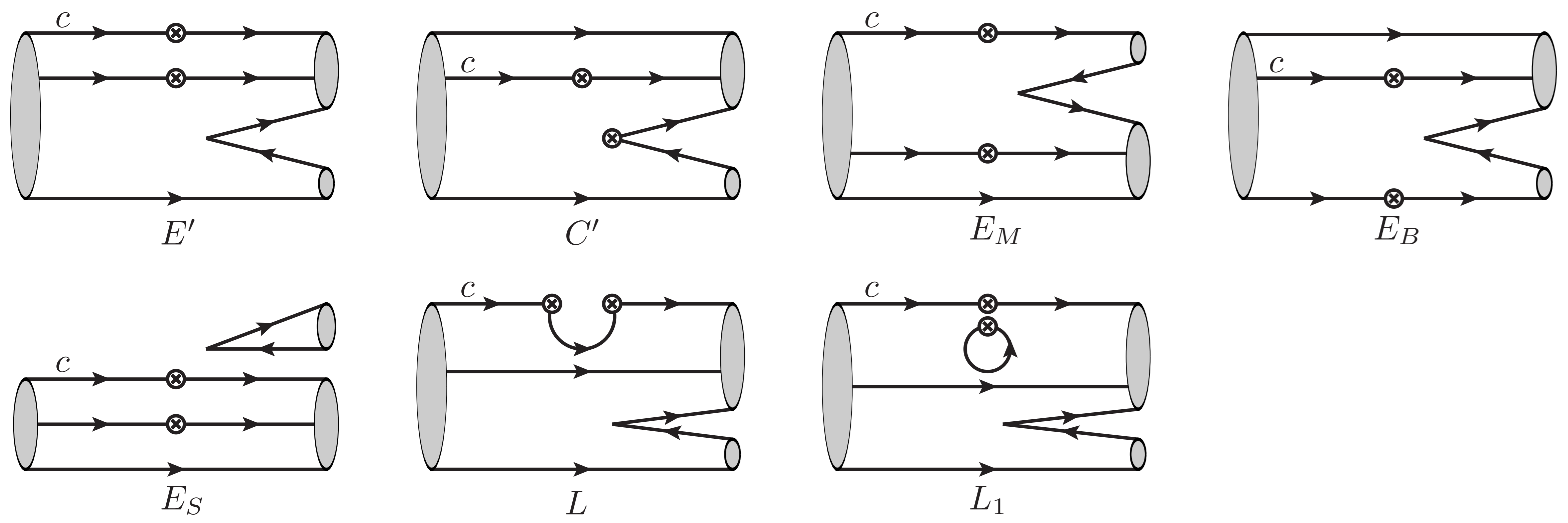}
  \caption{Topological diagrams in charmed anti-triplet baryon decays into a light meson and a light decuplet baryon.}\label{top1}
\end{figure}

If index contraction is understood as quark flow, each term in the Eq.~\eqref{am1} represents a topological amplitude.
The topological diagrams of the $\mathcal{B}_{c\overline 3}\to \mathcal{B}_{10}M$ decays are shown in Fig.~\ref{top1}.
The first five diagrams in Fig.~\ref{top1}, labeled $E^\prime$, $C^\prime$, $E_M$, $E_B$ and $E_S$, do not include quark-loop.
Diagrams $L$ and $L_1$ are quark-loop contributions.
By inserting tree and penguin operators into these diagrams, the tree and penguin induced amplitudes are obtained.
The amplitudes of $\mathcal{B}_{c\overline 3}\to \mathcal{B}_{10}M$ decays are listed in Table.~\ref{amp7}, where the penguin induced amplitudes are labeled with a superscript "$P$" to distinguish them from the tree induced amplitudes.
In the SM, the tree induced diagrams with $d$ quark loop and $s$ quark loop always appear simultaneously.
The penguin induced diagrams appear together with tree induced diagram $L$ with a fixed combination of $E^{\prime P}-C^{\prime P}-E^P_B-L^P-3L_1^{P}$.
In the flavor $SU(3)$ symmetry, the CKM matrix elements of the tree induced diagrams with quark loop are always $(V_{cd}^*V_{ud}+V_{cd}^*V_{ud})$.
Since $V_{cd}^*V_{ud}+V_{cs}^*V_{us}+V_{cb}^*V_{ub}=0$
and $|V_{cb}^*V_{ub}| \ll |V_{cd}^*V_{ud}|$ $\&$ $|V_{cs}^*V_{us}|$, all the penguin induced diagrams and tree induced diagrams with quark loop are negligible in branching fractions.
But they are not negligible in CP asymmetries due to the large weak phase in $V_{ub}$.

\begin{table}
\caption{Topological amplitudes of $\mathcal{B}_{c\overline 3}\to \mathcal{B}_{10}M$ decays, where the penguin induced amplitudes are labeled superscript "$P$" to distinguish from the tree induced amplitudes.}\label{amp7}
\begin{tabular}{|c|c|c|c|}
\hline\hline
 channel & amplitude & channel &amplitude\\\hline
 $\Lambda^+_c\to \Sigma^{*+}\pi^0$ & $\frac{1}{\sqrt{6}}\lambda_1(E^\prime-C^\prime+E_B)$ &$\Lambda^+_c\to \Delta^+K^0$ & $\frac{1}{\sqrt{3}}\lambda_2C^\prime$ \\\hline
 $\Lambda^+_c\to \Sigma^{*+}\eta_8$ & $\frac{1}{3\sqrt{2}}\lambda_1(E^\prime+C^\prime-2E_M+E_B)$ &$\Lambda^+_c\to \Delta^0K^+$ & $-\frac{1}{\sqrt{3}}\lambda_2C^\prime$ \\\hline
 $\Lambda^+_c\to \Sigma^{*+}\eta_1$ & ~~~~~~$\frac{1}{3}\lambda_1(E^\prime+C^\prime+E_M+E_B+3E_S)$~~~~~~ &$\Xi^+_c\to \Delta^+\eta_8$ & $\frac{1}{3\sqrt{2}}\lambda_2(E^\prime-2C^\prime+E_M+E_B)$ \\\hline
 $\Lambda^+_c\to \Sigma^{*0}\pi^+$ & $\frac{1}{\sqrt{6}}\lambda_1(E^\prime-C^\prime+E_B)$ & $\Xi^+_c\to \Delta^+\eta_1$& $\frac{1}{3}\lambda_2(E^\prime+C^\prime+E_M+E_B+3E_S)$ \\\hline
 $\Lambda^+_c\to \Delta^{++}K^-$ & $\lambda_1E_M$ & $\Xi^+_c\to \Sigma^{*0}K^+$& $\frac{1}{\sqrt{6}}\lambda_2(E^\prime-C^\prime+E_B)$ \\\hline
 $\Lambda^+_c\to \Delta^{+}\overline K^0$ & $\frac{1}{\sqrt{3}}\lambda_1E_M$ &$\Xi^+_c\to \Delta^{++}\pi^-$ & $\lambda_2E_M$ \\\hline
 $\Lambda^+_c\to \Xi^{* 0}K^+$ & $\frac{1}{\sqrt{3}}\lambda_1(E^\prime+E_B)$ &$\Xi^+_c\to \Sigma^{*+}K^0$ & $\frac{1}{\sqrt{3}}\lambda_2E_M$ \\\hline
 $\Xi^+_c\to \Sigma^{* +}\overline K^0$ & $\frac{1}{\sqrt{3}}\lambda_1C^\prime$ &$\Xi^+_c\to \Delta^0\pi^+$ &$\frac{1}{\sqrt{3}}\lambda_2(E^\prime+E_B)$  \\\hline
 $\Xi^+_c\to \Xi^{* 0}\pi^+$ & $-\frac{1}{\sqrt{3}}\lambda_1C^\prime$ &$\Xi^+_c\to \Delta^+\pi^0$ & $\frac{1}{\sqrt{6}}\lambda_2(E^\prime-E_M+E_B)$ \\\hline
 $\Xi^0_c\to \Sigma^{* 0}\overline K^0$ & $\frac{1}{\sqrt{6}}\lambda_1(-E^\prime+C^\prime-E_M)$ & $\Xi^0_c\to \Delta^0\eta_8$& $\frac{1}{3\sqrt{2}}\lambda_2(E^\prime-2C^\prime+E_M+E_B)$ \\\hline
 $\Xi^0_c\to \Xi^{* 0}\pi^0$ & $\frac{1}{\sqrt{6}}\lambda_1(C^\prime-E_B)$ & $\Xi^0_c\to \Delta^0\eta_1$& $\frac{1}{3}\lambda_2(E^\prime+C^\prime+E_M+E_B+3E_S)$ \\\hline
 $\Xi^0_c\to \Xi^{* 0}\eta_8$ & $\frac{1}{3\sqrt{2}}\lambda_1(2E^\prime-C^\prime+2E_M-E_B)$ &$\Xi^0_c\to \Sigma^{*0}K^0$ & $\frac{1}{\sqrt{6}}\lambda_2(E^\prime-C^\prime+E_M)$ \\\hline
 $\Xi^0_c\to \Xi^{* 0}\eta_1$ & $-\frac{1}{3}\lambda_1(E^\prime+C^\prime+E_M+E_B+3E_S)$ & $\Xi^0_c\to \Delta^+\pi^-$& $\frac{1}{\sqrt{3}}\lambda_2(E^\prime+E_M)$ \\\hline
 $\Xi^0_c\to \Sigma^{* +}K^-$ & $-\frac{1}{\sqrt{3}}\lambda_1(E^\prime+E_M)$ &$\Xi^0_c\to \Delta^-\pi^+$ & $\lambda_2E_B$ \\\hline
 $\Xi^0_c\to \Xi^{*-}\pi^+$ & $-\frac{1}{\sqrt{3}}\lambda_1E_B$ & $\Xi^0_c\to \Sigma^{*-}K^+$& $\frac{1}{\sqrt{3}}\lambda_2E_B$ \\\hline
 $\Xi^0_c\to \Omega^{ -}K^+$ & $-\lambda_1E_B$ &$\Xi^0_c\to \Delta^0\pi^0$ & $\frac{1}{\sqrt{6}}\lambda_2(-E^\prime-E_M+E_B)$ \\\toprule[1.2pt]
 $\Lambda^+_c\to \Delta^+\pi^0$&  \multicolumn{3}{c|}{$\frac{1}{\sqrt{6}}\lambda_d(E^\prime-C^\prime-E_M+E_B-2L)
 -\frac{2}{\sqrt{6}}\lambda_sL-\frac{2}{\sqrt{6}}\lambda_b(E^{\prime P}
 -C^{\prime P}-E^P_B-L^P-3L^P_1)$ } \\\hline
 $\Lambda^+_c\to \Delta^+\eta_8$&  \multicolumn{3}{c|}{ $\frac{1}{3\sqrt{2}}\lambda_d(E^\prime+C^\prime+E_M+E_B)$} \\\hline
 $\Lambda^+_c\to \Delta^+\eta_1$&  \multicolumn{3}{c|}{$\frac{1}{3}\lambda_d(E^\prime+C^\prime+E_M+E_B+3E_S)$ } \\\hline
 $\Lambda^+_c\to \Delta^0\pi^+$&  \multicolumn{3}{c|}{$\frac{1}{\sqrt{3}}\lambda_d(E^\prime-C^\prime+E_B-L)
 -\frac{1}{\sqrt{3}}\lambda_sL-\frac{1}{\sqrt{3}}\lambda_b(E^{\prime P}
 -C^{\prime P}-E^P_B-L^P-3L^P_1)$ } \\\hline
 $\Lambda^+_c\to \Sigma^{*+}K^0$&  \multicolumn{3}{c|}{$\frac{1}{\sqrt{3}}\lambda_d(E_M+L)
 +\frac{1}{\sqrt{3}}\lambda_s(C^\prime+L)-\frac{1}{\sqrt{3}}\lambda_b
 (-E^{\prime P}+C^{\prime P}+E^P_B+L^P+3L^P_1)$ } \\\hline
 $\Lambda^+_c\to \Sigma^{*0}K^+$&  \multicolumn{3}{c|}{$\frac{1}{\sqrt{6}}\lambda_d(E^\prime+E_B-L)
 -\frac{1}{\sqrt{6}}\lambda_s(C^\prime+L)-\frac{1}{\sqrt{6}}\lambda_b
 (E^{\prime P}-C^{\prime P}-E^P_B-L^P-3L^P_1)$ } \\\hline
 $\Lambda^+_c\to \Delta^{++}\pi^-$&  \multicolumn{3}{c|}{$\lambda_d(E_M+L)+\lambda_sL-\lambda_b
 (-E^{\prime P}+C^{\prime P}+E^P_B+L^P+3L^P_1)$ } \\\hline
 $\Xi^+_c\to \Delta^{+}\overline K^0$&  \multicolumn{3}{c|}{ $\frac{1}{\sqrt{3}}\lambda_d(C^\prime+L)+\frac{1}{\sqrt{3}}\lambda_s
 (E_M+L)-\frac{1}{\sqrt{3}}\lambda_b(-E^{\prime P}+C^{\prime P}+E^P_B+L^P+3L^P_1)$} \\\hline
 $\Xi^+_c\to \Sigma^{*0}\pi^+$&  \multicolumn{3}{c|}{ $-\frac{1}{\sqrt{6}}\lambda_d(C^\prime+L)+\frac{1}{\sqrt{6}}\lambda_s
 (E^\prime+E_B-L)-\frac{1}{\sqrt{6}}\lambda_b(E^{\prime P}-C^{\prime P}-E^P_B-L^P-3L^P_1)$} \\\hline
 $\Xi^+_c\to \Sigma^{*+}\eta_8$&  \multicolumn{3}{c|}{$-\frac{1}{\sqrt{2}}\lambda_dL
 +\frac{1}{3\sqrt{2}}\lambda_s(E^\prime-2C^\prime-2E_M+E_B-3L)
 -\frac{1}{\sqrt{2}}\lambda_b(E^{\prime P}-C^{\prime P}-E^P_B-L^P-3L^P_1)$ } \\\hline
 $\Xi^+_c\to \Sigma^{*+}\eta_1$&  \multicolumn{3}{c|}{ $\frac{1}{3}\lambda_s(E^\prime+C^\prime+E_M+E_B+3E_S)$} \\\hline
 $\Xi^+_c\to \Xi^{*0}K^+$&  \multicolumn{3}{c|}{ $-\frac{1}{\sqrt{3}}\lambda_dL
 +\frac{1}{\sqrt{3}}\lambda_s(E^\prime-C^\prime+E_B-L)
 -\frac{1}{\sqrt{3}}\lambda_b(E^{\prime P}-C^{\prime P}-E^P_B-L^P-3L^P_1)$} \\\hline
 $\Xi^+_c\to \Delta^{++}K^-$&  \multicolumn{3}{c|}{  $\lambda_dL+\lambda_s(E_M+L)-\lambda_b(-E^{\prime P}+C^{\prime P}+E^P_B+L^P+3L^P_1)$} \\\hline
 $\Xi^+_c\to \Sigma^{*+}\pi^0$&  \multicolumn{3}{c|}{ $-\frac{1}{\sqrt{6}}\lambda_dL
 +\frac{1}{\sqrt{6}}\lambda_s(E^\prime+E_B-L)
 -\frac{1}{\sqrt{6}}\lambda_b(E^{\prime P}-C^{\prime P}-E^P_B-L^P-3L^P_1)$} \\\hline
 $\Xi^0_c\to \Delta^{0}\overline K^0$&  \multicolumn{3}{c|}{ $\frac{1}{\sqrt{3}}\lambda_d(-E^\prime+C^\prime+L)+
 \frac{1}{\sqrt{3}}\lambda_s(E_M+L)-
 \frac{1}{\sqrt{3}}\lambda_b(-E^{\prime P}+C^{\prime P}+E^P_B+L^P+3L^P_1)$} \\\hline
 $\Xi^0_c\to \Sigma^{*0}\pi^0$&  \multicolumn{3}{c|}{$\frac{1}{2\sqrt{3}}\lambda_d(C^\prime+E_M-E_B+L)
 +\frac{1}{2\sqrt{3}}\lambda_s(-E^\prime+E_B+L)
 -\frac{1}{2\sqrt{3}}\lambda_b(-E^{\prime P}+C^{\prime P}+E^P_B+L^P+3L^P_1)$ } \\\hline
 $\Xi^0_c\to \Sigma^{*0}\eta_8$&  \multicolumn{3}{c|}{ $\frac{1}{6}\lambda_d(2E^\prime-C^\prime-E_M-E_B-3L)
 +\frac{1}{6}\lambda_s(E^\prime-2C^\prime-2E_M+E_B-3L)$} \\
 & \multicolumn{3}{c|}{$-\frac{1}{2}\lambda_b(E^{\prime P}-C^{\prime P}-E^P_B
 -L^P-3L^P_1)$}\\\hline
 $\Xi^0_c\to \Sigma^{*0}\eta_1$&  \multicolumn{3}{c|}{$\frac{1}{3\sqrt{2}}\lambda_d(-E^\prime-C^\prime-E_M-E_B-3E_S)
 +\frac{1}{3\sqrt{2}}\lambda_s(E^\prime+C^\prime+E_M+E_B+3E_S)$ } \\\hline
 $\Xi^0_c\to \Xi^{*0}K^0$&  \multicolumn{3}{c|}{ $-\frac{1}{\sqrt{3}}\lambda_d(E_M+L)
 +\frac{1}{\sqrt{3}}\lambda_s(E^\prime-C^\prime-L)
 -\frac{1}{\sqrt{3}}\lambda_b(E^{\prime P}-C^{\prime P}-E^P_B-L^P-3L^P_1)$} \\\hline
 $\Xi^0_c\to \Sigma^{*+}\pi^-$&  \multicolumn{3}{c|}{ $-\frac{1}{\sqrt{3}}\lambda_d(E_M+L)
 +\frac{1}{\sqrt{3}}\lambda_s(E^\prime-L)
 -\frac{1}{\sqrt{3}}\lambda_b(E^{\prime P}-C^{\prime P}-E^P_B-L^P-3L^P_1)$} \\\hline
 $\Xi^0_c\to \Sigma^{*-}\pi^+$&  \multicolumn{3}{c|}{ $-\frac{1}{\sqrt{3}}\lambda_dE_B
 +\frac{1}{\sqrt{3}}\lambda_sE_B$} \\\hline
 $\Xi^0_c\to \Xi^{*-}K^+$&  \multicolumn{3}{c|}{ $-\frac{1}{\sqrt{3}}\lambda_dE_B+\frac{1}{\sqrt{3}}\lambda_sE_B$} \\\hline
 $\Xi^0_c\to \Delta^{+}K^-$&  \multicolumn{3}{c|}{$-\frac{1}{\sqrt{3}}\lambda_d(E^\prime-L)
 +\frac{1}{\sqrt{3}}\lambda_s(E_M+L)
 -\frac{1}{\sqrt{3}}\lambda_b(-E^{\prime P}+C^{\prime P}+E^P_B+L^P+3L^P_1)$ } \\\hline
  \hline
\end{tabular}
\end{table}

The K\"orner-Pati-Woo theorem was used to reduce the amplitudes of charmed baryon decays in literature \cite{Geng:2019awr,Geng:2023pkr,Chau:1995gk,Hsiao:2020iwc,Wang:2024ztg,Zhao:2018mov,Groote:2021pxt,Hsiao:2021nsc,Zhong:2024zme,Zhong:2024qqs}, It states that if the two quarks produced by weak operators enter one baryon, they must be anti-symmetric in flavor \cite{Korner:1970xq,Pati:1970fg}.
In diagrams $E^\prime$, $C^\prime$, $E_M$, $E_B$, and $E_S$, the two quark lines emitted from weak vertex enter the final-state or resonance-state baryon.
Consequently, these quarks should be anti-symmetric in flavor according to the KPW theorem.
In the $SU(3)$ picture, the weak operator $\mathcal{O}^{k}_{ij}$ is decomposed as
\begin{align}
  \mathcal{O}^k_{ij}= &\frac{1}{8}\mathcal{O}(15)^k_{ij}+\frac{1}{4}\epsilon_{ijl}\mathcal{O}(\overline 6)^{lk}+\delta_j^k\Big(\frac{3}{8}\mathcal{O}( 3_t)_i-\frac{1}{8}\mathcal{O}(3_p)_i\Big)+
  \delta_i^k\Big(\frac{3}{8}\mathcal{O}( 3_p)_j-\frac{1}{8}\mathcal{O}( 3_t)_j\Big).
\end{align}
In the Standard Model, the 3-dimensional operators do not contribute to the diagrams without quark loops \cite{Wang:2024ztg}.
The two flavor indices $i, j$ in 6- and 15-dimensional operators are  $i\leftrightarrow j$ anti-symmetric and symmetric, respectively.
According to the above analysis, only $\mathcal{O}(\overline 6)$ operators contribute to diagrams $E^\prime$, $C^\prime$, $E_M$, $E_B$, and $E_S$.
The Fierz transformations of $\mathcal{O}(\overline 6)$ operators are opposite to the original ones, $\mathcal{F}[\mathcal{O}(\overline 6)] = -\mathcal{O}(\overline 6)$.
Thus, if two diagrams are connected by the Fierz transformation, they should be opposite according to the KPW theorem.
And if the Fierz transformation of a diagram is itself, it is forbidden by the KPW theorem.
The Fierz transformations of $E^\prime$, $C^\prime$, and $E_S$ are themselves. And the Fierz transformation of $E_M$ results in $E_B$, vice versa.
Consequently, we derive the following conclusions for topological diagrams,
\begin{align}\label{x1}
  E^\prime=C^\prime=E_S=0,\qquad E_M=-E_B.
\end{align}
Note that the equation $E_M=-E_B$ was not highlighted in literature such as Ref.~\cite{Hsiao:2020iwc}.

Eq.~\eqref{x1} indicates that only one independent amplitude contributes to branching fractions in the $SU(3)_F$ limit, which is consistent with Ref.~\cite{Geng:2019awr}.
Then we conclude that the dominant decay amplitudes of all $\mathcal{B}_{c\overline 3}\to \mathcal{B}_{10}M$ modes have definite proportional relations.
And the Lee-Yang parameters of all $\mathcal{B}_{c\overline 3}\to \mathcal{B}_{10}M$ decays, which describe the interference between $P$- and $D$-wave amplitudes, are the same in the $SU(3)_F$ limit.
According to Table.~\ref{amp7} and Eq.~\eqref{x1}, it is found the Cabibbo-favored modes $\Xi^+_c\to \Sigma^{* +}\overline K^0$ and $\Xi^+_c\to \Xi^{* 0}\pi^+$, the singly Cabibbo-suppressed modes $\Lambda^+_c\to \Delta^+\eta$ and $\Lambda^+_c\to \Delta^+\eta^\prime$, as well as the doubly Cabibbo suppressed modes $\Lambda^+_c\to \Delta^+K^0$, $\Lambda^+_c\to \Delta^0K^+$, $\Xi^+_c\to \Delta^+\eta$, $\Xi^+_c\to \Delta^+\eta^\prime$, $\Xi^0_c\to \Delta^0\eta$ and $\Xi^0_c\to \Delta^0\eta^\prime$ are forbidden by the KPW theorem in the $SU(3)_F$ limit.
Eq.~\eqref{x1} can be examined by global fitting of the $\mathcal{B}_{c\overline 3}\to \mathcal{B}_{10}M$ decays.
Actually, the global fitting has been performed in Ref.~\cite{Hsiao:2020iwc}.
The results show that
\begin{align}
|E_M/E_B|= 1.21\pm 0.14,\qquad \delta_{E_M}-\delta_{E_B}=(180.0\pm35.8)^\circ,
\end{align}
which are consistent with the KPW theorem predictions.

Considering the quark loop diagrams, there are two independent amplitudes, $E_M$ and $L^\prime = L-E^P_B-L^P-3L_1^{P}$, contribute to CP asymmetries in the $SU(3)_F$ limit.
If we define the ratio of $L^\prime$ and $E_M$ as
 $L^\prime/E_M = r e^{i\delta}$,
it is found there are only four possible values for the CP asymmetries of $\mathcal{B}_{c\overline 3}\to \mathcal{B}_{10}M$ decays in the $SU(3)_F$ limit:
\begin{align}
 A^{\pm}_{
 \rm CP}=\pm\frac{2|\lambda_b|r\sin\delta\sin\gamma}{|\lambda_s|},\qquad A^{1/3}_{\rm CP}=\frac{2|\lambda_b|r\sin\delta\sin\gamma}{3|\lambda_s|}, \qquad A^{0}_{\rm CP}=0.
\end{align}
Specifically, the direct CP asymmetries of singly Cabibbo-suppressed modes are
\begin{align}\label{a4}
 &A_{\rm CP}(\Lambda^+_c\to \Delta^+\pi^0)= A_{\rm CP}(\Lambda^+_c\to \Delta^0\pi^+)= A_{\rm CP}(\Lambda^+_c\to \Sigma^{*+}K^0)=A_{\rm CP}(\Lambda^+_c\to \Sigma^{*0}K^+)\nonumber\\&~~~~~~~~=A_{\rm CP}(\Lambda^+_c\to \Delta^{++}\pi^-)=A_{\rm CP}(\Xi^0_c\to \Xi^{*0}K^0)=A_{\rm CP}(\Xi^0_c\to \Sigma^{*+}\pi^-)=A^{+}_{\rm CP},
\end{align}
\begin{align}\label{a5}
&A_{\rm CP}(\Xi^+_c\to \Delta^{+}\overline K^0)= A_{\rm CP}(\Xi^+_c\to \Sigma^{*0}\pi^+)=A_{\rm CP}(\Xi^+_c\to \Sigma^{*+}\eta_8)\nonumber\\&~~~~~~~~=A_{\rm CP}(\Xi^+_c\to \Xi^{*0}K^+)=A_{\rm CP}(\Xi^+_c\to \Delta^{++}K^-)=A_{\rm CP}(\Xi^+_c\to \Sigma^{*+}\pi^0)\nonumber\\&~~~~~~~~~~~~~~~=A_{\rm CP}(\Xi^0_c\to \Delta^{0}\overline K^0)=A_{\rm CP}(\Xi^0_c\to \Sigma^{*0}\eta_8)=A_{\rm CP}(\Xi^0_c\to \Delta^{+}K^-)=A^{-}_{\rm CP},
\end{align}
\begin{align}\label{a6}
 A_{\rm CP}(\Xi^0_c\to \Sigma^{*0}\pi^0)=A^{1/3}_{\rm CP},
\end{align}
\begin{align}\label{a7}
&A_{\rm CP}(\Lambda^+_c\to \Delta^+\eta_8)= A_{\rm CP}(\Lambda^+_c\to \Delta^+\eta_1)=A_{\rm CP}(\Xi^+_c\to \Sigma^{*+}\eta_1)\nonumber\\&~~~~~~~~=A_{\rm CP}(\Xi^0_c\to \Sigma^{*0}\eta_1)=A_{\rm CP}(\Xi^0_c\to \Sigma^{*-}\pi^+)=A_{\rm CP}(\Xi^0_c\to \Xi^{*-}K^+)=A^{0}_{\rm CP}.
\end{align}
In experiments, we can combine several decay modes to enhance the possibility of observing CP violation in charmed baryon decays.

\section{Topological amplitudes beyond the $SU(3)_F$ symmetry}\label{be}
\begin{table}
\caption{Topological amplitudes contributing to the $\mathcal{B}_{c\overline 3}\to \mathcal{B}_{10}M$ decays in the linear $SU(3)$ breaking, where $\theta_C$ is the Cabibbo angle.}\label{ampx}
\begin{tabular}{|c|c|c|c|}
\hline\hline
 channel & amplitude & channel &amplitude\\\hline
 $\Lambda^+_c\to \Sigma^{*+}\pi^0$ & $-\frac{1}{\sqrt{6}}(E_M-E^{\prime(1)}+C^{\prime(1)}-E_B^{(1)})$ &$\Lambda^+_c\to \Delta^+K^0$ & $0$ \\\hline
 $\Lambda^+_c\to \Sigma^{*+}\eta_8$ & $-\frac{1}{3\sqrt{2}}(3E_M-E^{\prime(1)}-C^{\prime(1)}+2E_M^{(1)}+2E_M^{(3)}-E_B^{(1)})$ &$\Lambda^+_c\to \Delta^0K^+$ & $0$ \\\hline
 $\Lambda^+_c\to \Sigma^{*+}\eta_1$ & ~~~~~~$\frac{1}{3}(E^{\prime(1)}+C^{\prime(1)}+E_M^{(1)}+E_M^{(3)}+E_B^{(1)}+3E_S^{(1)})$~~~~~~ &$\Xi^+_c\to \Delta^+\eta_8$ & $0$ \\\hline
 $\Lambda^+_c\to \Sigma^{*0}\pi^+$ & $-\frac{1}{\sqrt{6}}(E_M-E^{\prime(1)}+C^{\prime(1)}-E_B^{(1)})$ & $\Xi^+_c\to \Delta^+\eta_1$& $0$ \\\hline
 $\Lambda^+_c\to \Delta^{++}K^-$ & $E_M+E_M^{(1)}$ & $\Xi^+_c\to \Sigma^{*0}K^+$& $\frac{1}{\sqrt{6}}\sin^2\theta_C(E_M+E_M^{(2)}+E_M^{(3)})$ \\\hline
 $\Lambda^+_c\to \Delta^{+}\overline K^0$ & $\frac{1}{\sqrt{3}}(E_M+E_M^{(1)})$ &$\Xi^+_c\to \Delta^{++}\pi^-$ & $-\sin^2\theta_C(E_M+E_M^{(2)})$ \\\hline
 $\Lambda^+_c\to \Xi^{* 0}K^+$ & $-\frac{1}{\sqrt{3}}(E_M-E^{\prime(1)}+E_M^{(3)}-E_B^{(1)})$ &$\Xi^+_c\to \Sigma^{*+}K^0$ & $-\frac{1}{\sqrt{3}}\sin^2\theta_C(E_M+E_M^{(2)}+E_M^{(3)})$ \\\hline
 $\Xi^+_c\to \Sigma^{* +}\overline K^0$ & $\frac{1}{\sqrt{3}}C^{\prime(1)}$ &$\Xi^+_c\to \Delta^0\pi^+$ &$\frac{1}{\sqrt{3}}\sin^2\theta_C(E_M+E_M^{(2)})$  \\\hline
 $\Xi^+_c\to \Xi^{* 0}\pi^+$ & $-\frac{1}{\sqrt{3}}C^{\prime(1)}$ &$\Xi^+_c\to \Delta^+\pi^0$ & $\sqrt{\frac{2}{3}}\sin^2\theta_C(E_M+E_M^{(2)})$ \\\hline
 $\Xi^0_c\to \Sigma^{* 0}\overline K^0$ & $-\frac{1}{\sqrt{6}}(E_M+E^{\prime(1)}-C^{\prime(1)}+E_M^{(1)}+E_M^{(4)})$ & $\Xi^0_c\to \Delta^0\eta_8$& $0$ \\\hline
 $\Xi^0_c\to \Xi^{* 0}\pi^0$ & $\frac{1}{\sqrt{6}}(E_M+C^{\prime(1)}+E_M^{(4)}-E_B^{(1)})$ & $\Xi^0_c\to \Delta^0\eta_1$& $0$ \\\hline
 $\Xi^0_c\to \Xi^{* 0}\eta_8$ & $\frac{1}{3\sqrt{2}}(3E_M+2E^{\prime(1)}-C^{\prime(1)}+2E_M^{(1)}+2E_M^{(3)}
 +3E_M^{(4)}-E_B^{(1)})$ &$\Xi^0_c\to \Sigma^{*0}K^0$ & $-\frac{1}{\sqrt{6}}\sin^2\theta_C(E_M+E_M^{(2)}+E_M^{(3)})$ \\\hline
 $\Xi^0_c\to \Xi^{* 0}\eta_1$ & $-\frac{1}{3}(E^{\prime(1)}+C^{\prime(1)}+E_M^{(1)}+E_M^{(3)}+E_B^{(1)}+3E_S^{(1)})$ & $\Xi^0_c\to \Delta^+\pi^-$& $-\frac{1}{\sqrt{3}}\sin^2\theta_C(E_M+E_M^{(2)})$ \\\hline
 $\Xi^0_c\to \Sigma^{* +}K^-$ & $-\frac{1}{\sqrt{3}}(E_M+E^{\prime(1)}+E_M^{(1)}+E_M^{(4)})$ &$\Xi^0_c\to \Delta^-\pi^+$ & $\sin^2\theta_C(E_M+E_M^{(2)})$ \\\hline
 $\Xi^0_c\to \Xi^{*-}\pi^+$ & $\frac{1}{\sqrt{3}}(E_M+E_M^{(4)}-E_B^{(1)})$ & $\Xi^0_c\to \Sigma^{*-}K^+$& $\frac{1}{\sqrt{3}}\sin^2\theta_C(E_M+E_M^{(2)}+E_M^{(3)})$ \\\hline
 $\Xi^0_c\to \Omega^{ -}K^+$ & $E_M+E_M^{(3)}+E_M^{(4)}-E_B^{(1)}$ &$\Xi^0_c\to \Delta^0\pi^0$ & $\sqrt{\frac{2}{3}}\sin^2\theta_C(E_M+E_M^{(2)})$ \\\toprule[1.2pt]
 $\Lambda^+_c\to \Delta^+\pi^0$&  \multicolumn{3}{c|}{$\sqrt{\frac{2}{3}}\sin\theta_C(E_M-L^{(1)})$ } \\\hline
 $\Lambda^+_c\to \Delta^+\eta_8$&  \multicolumn{3}{c|}{ $0$} \\\hline
 $\Lambda^+_c\to \Delta^+\eta_1$&  \multicolumn{3}{c|}{$0$ } \\\hline
 $\Lambda^+_c\to \Delta^0\pi^+$&  \multicolumn{3}{c|}{$\frac{1}{\sqrt{3}}\sin\theta_C(E_M-L^{(1)})$ } \\\hline
 $\Lambda^+_c\to \Sigma^{*+}K^0$&  \multicolumn{3}{c|}{$-\frac{1}{\sqrt{3}}\sin\theta_C(E_M-C^{\prime(1)}+E_M^{(3)}
 -L^{(1)})$ } \\\hline
 $\Lambda^+_c\to \Sigma^{*0}K^+$&  \multicolumn{3}{c|}{$\frac{1}{\sqrt{3}}\sin\theta_C(E_M-C^{\prime(1)}+E_M^{(3)}
 -L^{(1)})$ } \\\hline
 $\Lambda^+_c\to \Delta^{++}\pi^-$&  \multicolumn{3}{c|}{$-\sin\theta_C(E_M-L^{(1)})$ } \\\hline
 $\Xi^+_c\to \Delta^{+}\overline K^0$&  \multicolumn{3}{c|}{ $\frac{1}{\sqrt{3}}\sin\theta_C(E_M+E_M^{(1)}+E_M^{(2)}+L^{(1)})$} \\\hline
 $\Xi^+_c\to \Sigma^{*0}\pi^+$&  \multicolumn{3}{c|}{ $-\frac{1}{\sqrt{6}}\sin\theta_C(E_M-E^{\prime(1)}+E_M^{(2)}-E_B^{(1)}-L^{(1)})$} \\\hline
 $\Xi^+_c\to \Sigma^{*+}\eta_8$&  \multicolumn{3}{c|}{$
 -\frac{1}{3\sqrt{2}}\sin\theta_C(3E_M-E^{\prime(1)}+2C^{\prime(1)}+2E_M^{(1)}
 +3E_M^{(2)}+2E_M^{(3)}-E_B^{(1)}+3L^{(1)})$ } \\\hline
 $\Xi^+_c\to \Sigma^{*+}\eta_1$&  \multicolumn{3}{c|}{ $\frac{1}{3}\sin\theta_C(E^{\prime(1)}+C^{\prime(1)}+E_M^{(1)}+E_M^{(3)}+E_B^{(1)}
 +3E_S^{(1)})$} \\\hline
 $\Xi^+_c\to \Xi^{*0}K^+$&  \multicolumn{3}{c|}{ $-\frac{1}{\sqrt{3}}\sin\theta_C(E_M-E^{\prime(1)}+C^{\prime(1)}+E_M^{(2)}+E_M^{(3)}-E_B^{(1)}
 +L^{(1)})$} \\\hline
 $\Xi^+_c\to \Delta^{++}K^-$&  \multicolumn{3}{c|}{  $\sin\theta_C(E_M+E_M^{(1)}+E_M^{(2)}+L^{(1)})$} \\\hline
 $\Xi^+_c\to \Sigma^{*+}\pi^0$&  \multicolumn{3}{c|}{ $-\frac{1}{\sqrt{6}}\sin\theta_C(E_M-E^{\prime(1)}+E_M^{(2)}-E_B^{(1)}+L^{(1)})
 $} \\\hline
 $\Xi^0_c\to \Delta^{0}\overline K^0$&  \multicolumn{3}{c|}{ $
 \frac{1}{\sqrt{3}}\sin\theta_C(E_M+E_M^{(1)}+E_M^{(2)}+L^{(1)})$} \\\hline
 $\Xi^0_c\to \Sigma^{*0}\pi^0$&  \multicolumn{3}{c|}{$-\frac{1}{2\sqrt{3}}\sin\theta_C
 (3E_M+E^{\prime(1)}+E_M^{(2)}+2E_M^{(4)}-E_B^{(1)}-L^{(1)})$ } \\\hline
 $\Xi^0_c\to \Sigma^{*0}\eta_8$&  \multicolumn{3}{c|}{ $-\frac{1}{6}\sin\theta_C(3E_M-E^{\prime(1)}+2C^{\prime(1)}+2E_M^{(1)}+3E_M^{(2)}
 +2E_M^{(3)}-E_B^{(1)}+3L^{(1)})$}\\\hline
 $\Xi^0_c\to \Sigma^{*0}\eta_1$&  \multicolumn{3}{c|}{$\frac{1}{3\sqrt{2}}\sin\theta_C(E^{\prime(1)}+C^{\prime(1)}
 +E_M^{(1)}
 +E_M^{(3)}+E_B^{(1)}+3E_S^{(1)})$ } \\\hline
 $\Xi^0_c\to \Xi^{*0}K^0$&  \multicolumn{3}{c|}{ $\frac{1}{\sqrt{3}}\sin\theta_C(E_M+E^{\prime(1)}-C^{\prime(1)}+E_M^{(3)}+E_M^{(4)}
 -L^{(1)})$} \\\hline
 $\Xi^0_c\to \Sigma^{*+}\pi^-$&  \multicolumn{3}{c|}{ $\frac{1}{\sqrt{3}}\sin\theta_C(E_M+E^{\prime(1)}+E_M^{(4)}
 -L^{(1)})$} \\\hline
 $\Xi^0_c\to \Sigma^{*-}\pi^+$&  \multicolumn{3}{c|}{ $-\frac{1}{\sqrt{3}}\sin\theta_C(2E_M +E_M^{(2)}+E_M^{(4)}-E_B^{(1)}
)$} \\\hline
 $\Xi^0_c\to \Xi^{*-}K^+$&  \multicolumn{3}{c|}{ $-\frac{1}{\sqrt{3}}\sin\theta_C(2E_M+E_M^{(2)}+2E_M^{(3)}+E_M^{(4)}-E_B^{(1)})$} \\\hline
 $\Xi^0_c\to \Delta^{+}K^-$&  \multicolumn{3}{c|}{$\frac{1}{\sqrt{3}}\sin\theta_C(E_M+E_M^{(1)}+E_M^{(2)}+L^{(1)})$ } \\\hline
  \hline
\end{tabular}
\end{table}

The global fitting for the $\mathcal{B}_{c\overline 3}\to \mathcal{B}_{10}M$ modes in the $SU(3)_F$ limit is compromised by uncontrollable $SU(3)$ breaking effects.
In this section, we analyze the topological amplitudes of $\mathcal{B}_{c\overline 3}\to \mathcal{B}_{10}M$ decays in the linear $SU(3)_F$ breaking.
The decays of $D$ mesons into two pseudoscalar mesons were analyzed in the topological amplitudes with linear $SU(3)_F$ breaking in Ref.~\cite{Muller:2015lua}. Considering the first-order $SU(3)_F$ breaking, the total Hamiltonian is given as $\mathcal{H} = \mathcal{H}_0+\mathcal{H}_1$. Here, $\mathcal{H}_0$ is the QCD Hamiltonian with $m_u=m_d=m_s$, while $\mathcal{H}_1$ includes of the weak $|\Delta C|=1$ Hamiltonian $ \mathcal{H}_{ W}$ and the $SU(3)_F$ breaking Hamiltonian $\mathcal{H}_{\rm \cancel{SU(3)_F}} = (m_s-m_d)\overline ss$.
In Ref.~\cite{Wang:2020gmn}, the topological amplitudes with linear $SU(3)_F$ breaking were expressed in tensor form.
The same trick can be applied to $\mathcal{B}_{c\overline 3}\to \mathcal{B}_{10}M$ decays.
The decay amplitude of the $\mathcal{B}_{c\overline 3}\to \mathcal{B}_{10}M$ mode with linear $SU(3)_F$ breaking is constructed  as follows,
\begin{align}\label{am2}
  \mathcal{A}_{\rm \cancel{SU(3)_F}}(\mathcal{B}_{c\overline 3}\to \mathcal{B}_{10}M)& =
  E_M(\mathcal{B}_{c\overline3})_{ij}H^j_{kl}M^l_m \mathcal{B}_{10}^{ikm}
+  E_B(\mathcal{B}_{c\overline3})_{ij}H^j_{kl}M^k_m \mathcal{B}_{10}^{ilm}+E_M^{(1)}(\mathcal{B}_{c\overline3})_{ij}H^j_{kl}S^l_nM^n_m \mathcal{B}_{10}^{ikm}\nonumber\\&
+E_M^{(2)}(\mathcal{B}_{c\overline3})_{ij}H^n_{kl}S^j_nM^l_m \mathcal{B}_{10}^{ilm} +E_M^{(3)}(\mathcal{B}_{c\overline3})_{ij}H^j_{kl}S^m_nM^l_m \mathcal{B}_{10}^{ikn} +E_M^{(4)}(\mathcal{B}_{c\overline3})_{ij}H^j_{kl}S^i_nM^l_m \mathcal{B}_{10}^{nkm}\nonumber\\&+E_B^{(1)}(\mathcal{B}_{c\overline3})_{ij}H^j_{kl}S^l_nM^k_m\mathcal{B}_{10}^{inm}
+E_B^{(2)}(\mathcal{B}_{c\overline3})_{ij}H^n_{kl}S^j_nM^k_m \mathcal{B}_{10}^{ilm}+E_B^{(3)}(\mathcal{B}_{c\overline3})_{ij}H^j_{kl}S^m_nM^k_m \mathcal{B}_{10}^{iln}\nonumber\\&+E_B^{(4)}(\mathcal{B}_{c\overline3})_{ij}H^j_{kl}S^i_nM^k_m \mathcal{B}_{10}^{nlm}+E^{\prime(1)}(\mathcal{B}_{c\overline3})_{ij}H^j_{kl}S^l_nM^i_m \mathcal{B}_{10}^{knm}+C^{\prime(1)}(\mathcal{B}_{c\overline3})_{ij}H^k_{lm}S^m_nM^j_k \mathcal{B}_{10}^{iln}\nonumber\\&+E_S^{(1)}(\mathcal{B}_{c\overline3})_{ij}H^j_{kl}S^l_nM^m_m \mathcal{B}_{10}^{ikn}+L^{(1)}(\mathcal{B}_{c\overline3})_{ij}H^n_{kl}S^l_nM^j_m \mathcal{B}_{10}^{ikm}.
\end{align}
In Eq.~\eqref{am2}, the second-rank tensor $S$ is used to mark the $SU(3)_F$ breaking induced by $s$ quark mass,
\begin{align}
  S = \left(
        \begin{array}{ccc}
          0 & 0 & 0 \\
          0 & 0 & 0 \\
          0 & 0 & 1 \\
        \end{array}
      \right).
\end{align}
The decay amplitude with superscript "(i)" denotes the difference between two diagrams, one with and one without the $s$ quark.
For instance, $E_M^{(1)}$ represents the difference between two $E_M$ diagrams with $l = n = s$ and $l,n = u$ or $d$.
The emergence of new topologies is similar to the splitting of energy level in atoms.
When the $SU(3)_F$ symmetry breaks down to $SU(2)_F$ symmetry, the original diagram splits into several diagrams.

\begin{figure}
  \centering
  \includegraphics[width=15cm]{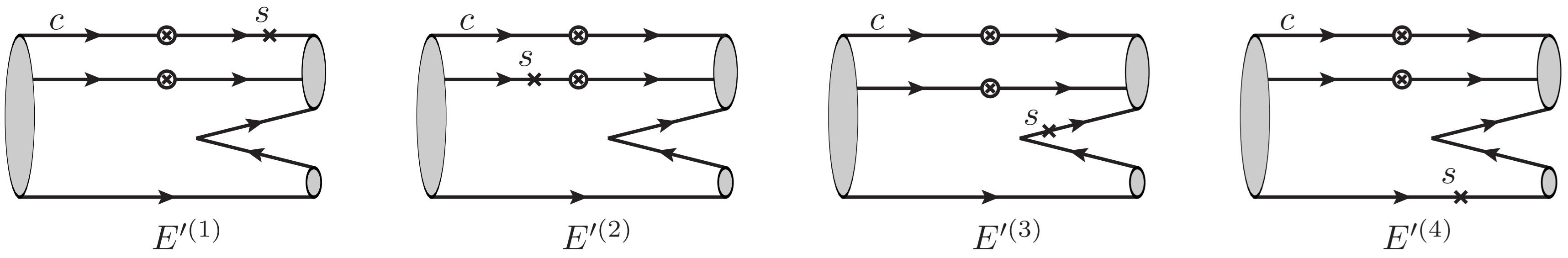}
  \caption{Topological diagrams $E^{\prime(1)}$, $E^{\prime(2)}$, $E^{\prime(3)}$ and $E^{\prime(4)}$, where "$\times$" is the $SU(3)_F$ breaking induced by $s$ quark mass.}\label{top2}
\end{figure}
The topological amplitudes disappearing in Eq.~\eqref{am2} are either forbidden by the KPW theorem or suppressed by small CKM matrix elements.
And the decay amplitudes defined in Eq.~\eqref{am2} have following relations according to the KPW theorem,
\begin{align}
  E_{M}=-E_{B},\qquad E_{M}^{(2)}=-E_{B}^{(2)}\qquad E_{M}^{(3)}=-E_{B}^{(3)},\qquad E_{M}^{(4)}=-E_{B}^{(4)}.
\end{align}
Taking $E^{\prime(i)}$ diagrams as examples, we explain above arguments as follow.
The topological diagrams $E^{\prime(1)}$, $E^{\prime(2)}$, $E^{\prime(3)}$ and $E^{\prime(4)}$ are shown in Fig.~\ref{top2}, in which "$\times$" is used to mark the $SU(3)_F$ breaking induced by $s$ quark mass.
One can find the flavor symmetry of the two quarks produced by weak vertex in $E^{\prime(2)}$, $E^{\prime(3)}$ and $E^{\prime(4)}$ is not broken by $s$ quark mass.
Thus diagrams $E^{\prime(2)}$, $E^{\prime(3)}$ and $E^{\prime(4)}$ are suppressed by the KPW theorem even considering the linear $SU(3)_F$ breaking.
Only $E^{\prime(1)}$ contributes to the $\mathcal{B}_{c\overline 3}\to \mathcal{B}_{10}M$ decays.

The amplitudes for the $\mathcal{B}_{c\overline 3}\to \mathcal{B}_{10}M$ decays in the linear $SU(3)_F$ breaking are listed in Table.~\ref{ampx}.
There are 10 amplitudes contributing to the $\mathcal{B}_{c\overline 3}\to \mathcal{B}_{10}M$ decays: $E_M$, $E^{\prime(1)}$, $C^{\prime(1)}$, $E_M^{(1)}$, $E_M^{(2)}$, $E_M^{(3)}$, $E_M^{(4)}$, $E_B^{(1)}$, $E_S^{(1)}$ and $L^{(1)}$ with $2\times 10 -1 = 19$ free parameters. We can fix the 19 free parameters by fitting branching fractions.
If the different partial wave contributions are considered, there are $2\times 2\times 10 -1 = 39$ free parameters.
The Lee-Yang parameters will be used to fix the 39 free parameters.
Compared to other decay modes, the advantage of $\mathcal{B}_{c\overline 3}\to \mathcal{B}_{10}M$ decay is that we have enough available data to extract the decay amplitudes in the linear $SU(3)$ breaking.
In other decay modes such as $D$ meson decays into two pseudoscalar mesons, the amount of free parameters in the linear $SU(3)_F$ breaking is larger than the available data, resulting in sizable uncertainties in the global fitting \cite{Muller:2015lua}.
Thereby, the $\mathcal{B}_{c\overline 3}\to \mathcal{B}_{10}M$ decays are ideal platforms to study $SU(3)$ breaking effects.

Since the available data are sufficient to fix the free parameters in the linear $SU(3)$ breaking, we can extract the $SU(3)_F$ breaking part of the quark loop diagram, $L^{(1)} = L^{s} - L^{d}$, by global fitting of the branching fractions of $\mathcal{B}_{c\overline 3}\to \mathcal{B}_{10}M$ decays.
This will allow us to analyze the potential size the penguin diagram in charm sector and gain a better understanding of the large CP violation observed in charm meson decays \cite{LHCb:2022lry,Aaij:2019kcg}.
We suggest the measure the branching fractions of the Cabibbo-favored and the singly Cabibo-suppressed $\mathcal{B}_{c\overline 3}\to \mathcal{B}_{10}M$ modes as many as possible.
It will provide further insight into the quark-loop diagram in the charm sector.

From Table.~\ref{ampx}, it is found the Cabibbo-favored modes $\Xi^+_c\to \Sigma^{* +}\overline K^0$ and $\Xi^+_c\to \Xi^{* 0}\pi^+$ are not forbidden by the KPW theorem after including the linear $SU(3)_F$ breaking effects, which is beyond the theoretical predictions in the $SU(3)_F$ limit \cite{Geng:2017mxn,Sharma:1996sc,Korner:1992wi,Geng:2019awr,Xu:1992sw,Hsiao:2020iwc}.
In consideration of the $SU(3)_F$ breaking effects, it is possible the branching fractions of these two modes are at order of $\mathcal{O}(10^{-4})$ or even $\mathcal{O}(10^{-3})$.
As suggested in Ref.~\cite{Geng:2022yxb}, branching fractions of the $\Xi^+_c\to \Sigma^{* +}\overline K^0$ and $\Xi^+_c\to \Xi^{* 0}\pi^+$ modes could be used to test $SU(3)_F$ breaking effects.
Moreover, the decay amplitudes of the $\Xi^+_c\to \Sigma^{* +}\overline K^0$ and $\Xi^+_c\to \Xi^{* 0}\pi^+$ modes are identical in the first order $SU(3)_F$ breaking.
The difference between $\mathcal{A}(\Xi^+_c\to \Sigma^{* +}\overline K^0)$ and $\mathcal{A}(\Xi^+_c\to \Xi^{* 0}\pi^+)$ will indicate the second order $SU(3)_F$ breaking effects.
The branching fractions of the $\Xi^+_c\to \Sigma^{* +}\overline K^0$ and $\Xi^+_c\to \Xi^{* 0}\pi^+$ are given by \cite{PDG}
\begin{align}
\mathcal{B}r(\Xi^+_c\to \Sigma^{* +}\overline K^0) = (2.9\pm 2.0)\% ,\qquad
\mathcal{B}r(\Xi^+_c\to \Xi^{* 0}\pi^+) < 2.9\times 10^{-3}.
\end{align}
More precise measurements are desirable to study $SU(3)_F$ breaking effects.
On the other hand, the singly Cabibbo-suppressed modes $\Lambda^+_c\to \Delta^+\eta$, $\Lambda^+_c\to \Delta^+\eta^\prime$, and the doubly Cabibbo suppressed modes $\Lambda^+_c\to \Delta^+K^0$, $\Lambda^+_c\to \Delta^0K^+$, $\Xi^+_c\to \Delta^+\eta$, $\Xi^+_c\to \Delta^+\eta^\prime$, $\Xi^0_c\to \Delta^0\eta$, $\Xi^0_c\to \Delta^0\eta^\prime$ are still forbidden by the KPW theorem in the linear $SU(3)_F$ breaking.
Testing the reliability of the KPW theorem can be done by measuring the branching fractions of these forbidden modes.

The isospin symmetry is the most precise flavor symmetry.
The isospin breaking is naively predicted to be $\mathcal{O}(1\%)$.
Testing the KPW theorem with isospin symmetry in charmed baryon decays was
first proposed in Ref.~\cite{Geng:2019xbo} with the $\Lambda_c^+\to \Sigma^0K^+$ and $\Lambda_c^+\to \Sigma^+K^0$ decays, and was verified by the BESIII Collaboration \cite{BESIII:2022wxj}.
Due to the well-defined flavor symmetry of decuplet baryons, we can find more decay channels to test the KPW theorem under the isospin symmetry in the $\mathcal{B}_{c\overline 3}\to \mathcal{B}_{10}M$ decays.
The isospin sum rules for the $\mathcal{B}_{c\overline 3}\to \mathcal{B}_{10}M$ decays are derived in Refs.~\cite{Luo:2023vbx,Jia:2019zxi}.
We tested the isospin sum rule for the $\Lambda_c^+\to\Sigma^{*+}\pi^0$ and $\Lambda_c^+\to\Sigma^{*0}\pi^+$ modes in \cite{Luo:2023vbx}, in which the experimental data are consistent with the predictions of isospin symmetry.
According to Table.~\ref{ampx}, we find following isospin relations beyond the general isospin sum rules:
\begin{align}\label{ax1}
\sqrt{\frac{3}{2}}\mathcal{A}(\Lambda^+_c\to \Delta^{+}\pi^0) = \sqrt{3}\mathcal{A}(\Lambda^+_c\to \Delta^{0}\pi^+)=-\mathcal{A}(\Lambda^+_c\to \Delta^{++}\pi^-),
\end{align}
\begin{align}
\mathcal{A}(\Lambda^+_c\to \Sigma^{*+} K^0) = -\sqrt{2}\mathcal{A}(\Lambda^+_c\to \Sigma^{*0}K^+),
\end{align}
\begin{align}\label{ax2}
\mathcal{A}(\Xi^+_c\to \Delta^{++} K^-) = \sqrt{3}\mathcal{A}(\Xi^+_c\to \Delta^{+} \overline K^0)=\sqrt{3}\mathcal{A}(\Xi^0_c\to \Delta^{0} \overline K^0)=\sqrt{3}\mathcal{A}(\Xi^0_c\to \Delta^{+} K^-),
\end{align}
\begin{align}
\mathcal{A}(\Xi^+_c\to \Sigma^{*0}\pi^+) = \mathcal{A}(\Xi^+_c\to \Sigma^{*+}\pi^0),
\end{align}
\begin{align}
 \sqrt{2}\mathcal{A}(\Xi^+_c\to \Sigma^{*0} K^+) = \mathcal{A}(\Xi^0_c\to \Sigma^{*-} K^+),
\end{align}
\begin{align}
\mathcal{A}(\Xi^+_c\to \Sigma^{*+} K^0)=\sqrt{2}\mathcal{A}(\Xi^0_c\to \Sigma^{*0} K^0),
\end{align}
\begin{align}
&\mathcal{A}(\Xi^+_c\to \Delta^{++}\pi^-) = \sqrt{3}\mathcal{A}(\Xi^0_c\to \Delta^{+}\pi^-),
\end{align}
\begin{align}
&\sqrt{3}\mathcal{A}(\Xi^+_c\to \Delta^{0}\pi^+) = \mathcal{A}(\Xi^0_c\to \Delta^{-}\pi^+),
\end{align}
\begin{align}
&\mathcal{A}(\Xi^+_c\to \Delta^{+}\pi^-) = \mathcal{A}(\Xi^0_c\to \Delta^{0}\pi^0).
\end{align}
These equations hold after considering the quark-loop diagrams and $U$- and $V$-spin breaking effects, but they break down if the KPW theorem does not work.
One can test the KPW theorem under isospin symmetry by checking above equations through measuring branching fractions.
Besides, Eqs.~\eqref{ax1} $\sim$ \eqref{ax2} indicate the following CP violating relations in the isospin symmetry,
\begin{align}
A_{\rm CP}(\Lambda^+_c\to \Delta^{+}\pi^0) =A_{\rm CP}(\Lambda^+_c\to \Delta^{0}\pi^+)=A_{\rm CP}(\Lambda^+_c\to \Delta^{++}\pi^-),
\end{align}
\begin{align}
A_{\rm CP}(\Lambda^+_c\to \Sigma^{*+} K^0) = A_{\rm CP}(\Lambda^+_c\to \Sigma^{*0}K^+),
\end{align}
\begin{align}
A_{\rm CP}(\Xi^+_c\to \Delta^{++} K^-) = A_{\rm CP}(\Xi^+_c\to \Delta^{+} \overline K^0)=A_{\rm CP}(\Xi^0_c\to \Delta^{0} \overline K^0)=A_{\rm CP}(\Xi^0_c\to \Delta^{+} K^-).
\end{align}
These equations could help experiments select appropriate channels to search for CP asymmetries.

\section{Summary}\label{sum}
In this work, we first study the topological amplitudes of the $\mathcal{B}_{c\overline 3}\to \mathcal{B}_{10}M$ decays in the flavor $SU(3)$ limit.
Notice that most of topological diagrams are suppressed by the K\"orner-Pati-Woo theorem, we find only two independent amplitudes contribute to the $\mathcal{B}_{c\overline 3}\to \mathcal{B}_{10}M$ decays, with one of them  dominating the branching fractions.
Many decay channels are forbidden by the K\"orner-Pati-Woo theorem.
The dominated amplitudes of different $\mathcal{B}_{c\overline 3}\to \mathcal{B}_{10}M$ decay channels exhibit definite proportional relations.
The Lee-Yang parameters are the same in all $\mathcal{B}_{c\overline 3}\to \mathcal{B}_{10}M$ decays.
Moreover, there are only four possible values for the CP asymmetries of $\mathcal{B}_{c\overline 3}\to \mathcal{B}_{10}M$ decays.

The topological diagrams of the $\mathcal{B}_{c\overline 3}\to \mathcal{B}_{10}M$ decays in the linear $SU(3)_F$ breaking are also investigated.
We find there are ten topological amplitudes contributing to the $\mathcal{B}_{c\overline 3}\to \mathcal{B}_{10}M$ decays in the linear $SU(3)_F$ breaking.
The $SU(3)_F$ breaking part of the quark loop diagram $L^{(1)}$ can be extracted by a global fitting of branching fractions.
This could help us understand CP violation in the charm sector.
The Cabibbo-favored modes $\Xi^+_c\to \Sigma^{* +}\overline K^0$ and $\Xi^+_c\to \Xi^{* 0}\pi^+$, which are forbidden by the K\"orner-Pati-Woo theorem in the $SU(3)_F$ limit, have non-zero branching fractions after including the linear $SU(3)_F$ breaking effects.
Additionally, some isospin equations that are beyond the usual isospin sum rules are proposed to test the K\"orner-Pati-Woo theorem.

\begin{acknowledgements}

This work was supported in part by the National Natural Science Foundation of China under Grants No. 12105099.

\end{acknowledgements}

\end{document}